\documentclass[floatfix,superscriptaddress,amsmath,amssymb,twocolumn]{revtex4}

\usepackage{graphicx}

\newcommand{\intd}{\mathrm{d}}
\newcommand{\parref}[1]{(\ref{#1})}
\newcommand{\vect}[1]{\mathbf{#1}}

\newcommand{\matelemdirect}[3]{\left\langle #1 \left| #2 \right| #3 \right\rangle}

\newcommand{\expvaluedirect}[1]{\langle #1 \rangle}
\providecommand{\abs}[1]{\left|#1\right|}
\newcommand{\absdirect}[1]{|#1|}
\DeclareMathOperator{\csch}{csch}

\begin{document}

\title{Must Kohn-Sham Oscillator Strengths be Accurate at Threshold?}
\author{Zeng-hui Yang}
\affiliation{Department of Chemistry, University of California - Irvine, CA 92697, USA}
\author{Meta van Faassen}
\affiliation{Afdeling Theoretische Chemie, Vrije Universiteit, De Boelelaan 1083, 1081 HV Amsterdam, Kingdom of the 
Netherlands}
\author{Kieron Burke}
\affiliation{Department of Chemistry, University of California - Irvine, CA 92697, USA}
\date{\today}

\begin{abstract}
The exact ground-state Kohn-Sham(KS) potential for the helium atom is known from accurate wavefunction calculations 
of the ground-state density. The threshold for photoabsorption from this potential matches the physical system exactly. 
By carefully studying its absorption spectrum, we show the answer to the title question is no. To address this problem in 
detail, we generate a highly accurate simple fit of a two-electron spectrum near the threshold, and apply the method to 
both the experimental spectrum and that of the exact ground-state Kohn-Sham potential.
\end{abstract}

\maketitle

\section{Introduction}
\label{sect:intro}
Ground-state density-functional theory(DFT)\cite{HK64,KS65,FNM03,RCFB08} is enjoying more and more popularity for 
calculating various atomic and molecular properties. The balance between accuracy and calculation speed in DFT is 
achieved using an auxiliary Kohn-Sham(KS) system of non-interacting electrons. If the exact exchange-correlation 
energy were known as a functional of the density, DFT would yield exact ground-state energies.

In principle, all atomic and molecular properties are functionals of the ground-state density, including the properties of 
excited states\cite{DG90}, but in practice only the ground-state energy functional has been usefully approximated. The 
excited-state properties of the non-interacting Kohn-Sham reference system are often used to understand and even 
approximate those of the true interacting system, but in most cases this has no theoretical justification.  Thus the results 
of excited-state calculations with ground-state DFT must be carefully examined, since the KS orbitals and energies are 
(within ground-state DFT) artificial constructs designed only to reproduce the ground-state density. The more we 
understand about the differences between the KS system and the real system, the better we can determine whether an 
excited property of the KS system can be justified as an approximation to the real property.  We study the exactness of 
the KS oscillator strength at the first ionization threshold in this paper.

On the other hand, time-dependent density-functional theory(TDDFT) in principle gives several exact properties of 
excited states\cite{RG84}. Linear response TDDFT is a method that begins from ground-state DFT, and couples ground-
state KS transitions to give the correct properties of excited states.\cite{C96,PGG96} If we could use the exact time-
dependent functional, the TDDFT method would exactly generate the properties of the real system from the results of 
the ground-state KS calculation of systems with non-interacting electrons. Thus our study of the exactness of the KS 
oscillator strength is converted to a question about the difference betweeen ground-state DFT and TDDFT. 

This may seem to be a simple problem, since ground-state DFT is not designed to give the correct oscillator strength at 
the ionization threshold. The oscillator strengths can be extracted from strength of the poles of the linear response 
function, and the KS linear response function does not involve the Hartree-exchange-correlation(HXC) kernel(Refer to 
Eqn. \parref{eqn:bkgnd:dysonlike} and \parref{eqn:bkgnd:fhxc}), so there is no {\em a priori} reason to expect the exact 
KS system to  give the correct oscillator strength at the ionization threshold. By 'exact KS' we mean the KS potential as 
extracted from an extremely accurate ground-state density\cite{UG94}, thereby avoiding the difficulty of distinguishing 
the effect of approximate ground-state XC functionals from that of KS-DFT itself. However, the ionization threshold of the 
exact KS system {\em is} equal to the ionization threshold of the real system, since Koopmans' theorem holds exactly for 
the exact KS system.\cite{FNM03} Thus this specific excited-state property is given exactly by the KS system, despite the 
lack of input from the Hartree and XC kernels. This is the only known direct link between real excited-state properties 
and their KS counterparts, and such links have proven invaluable in studying and understanding both ground-state and 
TDDFT\cite{EFB09}.

Given the usefulness of such links, and how both the strength and position of the the threshold occur at the same 
frequency, it is important to ask the title question, to see if some unknown exact condition might be lurking beneath the 
surface.   To do this, we study one specific case.  If, for He, we find definitively that the threshold oscillator strength is not 
given by the KS system, the answer is definitely no, and this cannot be true in general. If we did find it to match, the title 
question would remain open, and we would look for other cases and/or a proof of the equality.  As we show below, the 
answer is indeed no.

This has important consequences for the unknown exact XC kernel of TDDFT. To shift an ionization threshold, the 
kernel would need to be complex, with a branch cut at the position of the KS threshold.  This is not the case for the first 
ionization threshold, but is for all higher ionizations.  On the other hand, since we show that, typically, the oscillator 
strength of the KS system {\em is} corrected by TDDFT, this means that, at the threshold, the Hartree-XC kernel {\em 
must} have some non-zero off-diagonal matrix elements.  To understand this, \cite{AGB03} showed that, in the absence 
of off-diagonal matrix elements, the KS oscillator strengths are unchanged by the action of the kernel.

\begin{figure}[htbp]
\centering
\includegraphics[angle=-90,keepaspectratio]{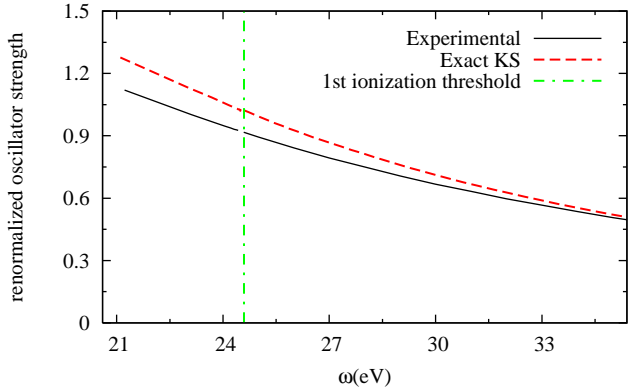}
\caption{KS and experimental single-electron oscillator strength of He near threshold.\cite{SHYH94,WMB03} The 
ionization threshold is at 0.9036 Hartree. The bound-region spectrum is renormalized with the factor $1/n_f^3$\,, where 
$n_f$ is the principal quantum number of the final state.}
\label{fig:thres}
\end{figure}

In this paper we will show the answer to the title question is no. To prove that the KS oscillator strength doesn't have to 
be exact at the ionization threshold, we only need to provide a counter-example, so we study the helium oscillator 
strength spectrum in this paper. The helium atom is the simplest multi-electron system, and thus a theorists' favorite. Fig. 
\ref{fig:thres} shows the photoabsorption spectrum of Helium near the ionization threshold(24.6 eV), and the two curves 
are of the real helium and of the exact ground-state KS helium. Fig. \ref{fig:thres} suggests the answer to the title 
question is no, but there could conceivably be near-degeneracies near the ionization threshold, and we wish to 
demonstrate that the oscillator strength curve can be expected to be smooth near the ionization threshold explicitly. 
Hence we use a fit to explicitly show that the oscillator strength curves of the real helium and the KS helium are smooth 
across the ionization threshold, showing there are no near-degeneracies at the threshold. This allows the comparison 
of the value of the oscillator strengths at the ionization threshold of these two systems, showing that the difference 
between the spectra near the threshold of KS and real systems is inherent. Since our purpose is to understand the 
difference between KS system and the real system, the fit is not done to the data points but to the general properties, 
such as the oscillator strength sum rules. We test our fit  on the hydrogen oscillator strength spectrum, and then apply 
the fit to KS helium, real helium, and the result of using  approximated TDDFT on the exact ground-state KS spectrum.

\section{Background}
\label{sect:bkgnd}
In this section we provide a brief definition of notation and concepts used in this article. For any interacting electronic 
problem, the exact ground-state KS system is described by the KS equations\cite{KS65}:
\begin{gather}
\label{eqn:bkgnd:KS}
\left\{-\frac{1}{2}\nabla^2+v_s(\vect{r})\right\}\psi_i(\vect{r})=\epsilon_i\psi_i(\vect{r}),\\
\intertext{where}
n(\vect{r})=\sum_{i=1}^N\abs{\psi_i(\vect{r})}^2.
\end{gather}
Here and unless otherwise noted, we use atomic units so that energies are in Hartrees and distances in Bohr radii. The 
KS orbitals and eigenenergies are denoted as $\psi_i(\vect{r})$ and $\epsilon_i$, and $v_s(\vect{r})$ is the KS potential, 
which can be separated into 3 pieces: external, Hartree, and exchange-correlation(XC) potential\cite{KS65, FNM03}. 
The exact dependence of the XC contribution on the density is unknown and many approximation schemes are 
available, but in this article we use the exact value\cite{UG94}. This is calculated by first obtaining the accurate density 
from a quantum Monte-Carlo calculation, then inserting the density into the KS equations and finding the potential that 
gives this density.\cite{UG94}

The absorption spectrum in terms of photoabsorption cross section $\sigma$ is defined as below\cite{F06}:
\begin{equation}
\label{eqn:bkgnd:specdef}
\sigma(\omega)=\frac{2\pi^2}{c}\sum_q f_q\delta(\omega-\omega_q)+\sigma_{\text{cont}}(\omega),
\end{equation}
where q denotes bound-to-bound transitions from state i to state f, $f_q$ is the oscillator strength of transition q, and $
\sigma_{\text{cont}}$ which begins at $\omega=I$ is the spectrum of the continuum region. For bound-to-bound 
transitions, the oscillator strengths are defined as:
\begin{equation}
\label{eqn:bkgnd:fdef}
f_{q}=2\omega_{q}\abs{\matelemdirect{\Psi_f}{\vect{\hat{r}}}{\Psi_i}}^2.
\end{equation}
As defined in Eqn. \parref{eqn:bkgnd:specdef}, the spectrum comprises the discrete bound-to-bound transitions and the 
continuous bound-to-continuum transitions. We define $\tilde{\sigma}(\omega)$ as the analytical continuation of $
\sigma_{\text{cont}}(\omega)$ for $\omega<I$. This can be found easily by considering the bound oscillator strength as 
a continous function of $\omega_q$, yielding \cite{F06}
\begin{equation}
\label{eqn:bkgnd:renorm}
\tilde{\sigma}(\omega)=\left.f(\omega)/\left(\frac{d\epsilon}{dn}\right)\right|_{\epsilon=\omega-I}.
\end{equation}
In reverse, the usual oscillator strength for transition $q=1s\to np$ is given by
\begin{equation}
\label{eqn:bkgnd:revnorm}
f_q=\left(\frac{d\epsilon}{dn}\right)\tilde{\sigma}(I+\epsilon),
\end{equation}
where $\epsilon$ is the energy of the np state.

The oscillator strengths are related to the dynamic polarizability  $\alpha(\omega)$ by the following equations:
\begin{equation}
\label{eqn:bkgnd:polarizability}
\alpha(\omega)=\int \intd^3r\int \intd^3r'\:z z'\tilde{\chi}(\vect{r},\vect{r}';\omega),
\end{equation}
\begin{equation}
\label{eqn:bkgnd:falpha}
\tilde{\sigma}(\omega)=\frac{4\pi\omega}{c}\Im[\alpha(\omega)],
\end{equation}
where $\chi(\vect{r},\vect{r}';\omega)$ is the linear response function of the real system, defined by the Fourier transform 
of the linear response function in time:
\begin{equation}
\label{eqn:bkgnd:respfunc}
\chi(\vect{r},t,\vect{r}',t')=\left.\frac{\delta n[v_\text{ext}](\vect{r}t)}{\delta v_\text{ext}(\vect{r}'t')}\right|_{v_\text{ext}[n_0]}.
\end{equation}
The polarizability and the linear response function of the KS spectrum is defined similarly. The KS linear response 
function is related to the real(or 'exact TDDFT') linear response function by a Dyson-like equation\cite{GK85}:
\begin{multline}
\label{eqn:bkgnd:dysonlike}
\tilde{\chi}(\vect{r},\vect{r}';\omega)=\tilde{\chi}_\text{KS}(\vect{r},\vect{r}';\omega)+\int\intd^3r_1\int\intd^3r_2\;\tilde{\chi}_
\text{KS}(\vect{r},\vect{r}_1;\omega)\\
\times\tilde{f}_\text{Hxc}(\vect{r}_1,\vect{r}_2;\omega)\tilde{\chi}(\vect{r}_2,\vect{r}';\omega),
\end{multline}
where $\tilde{f}_\text{Hxc}(\vect{r},\vect{r}';\omega)$ is the HXC kernel in frequency domain, defined as the Fourier 
transform of the HXC kernel in time domain:
\begin{equation}
\label{eqn:bkgnd:fhxc}
f_\text{Hxc}(\vect{r},\vect{r}';t-t')=\frac{1}{\abs{\vect{r}-\vect{r}'}}+\frac{\delta v_\text{xc}(\vect{r},t)}{\delta n(\vect{r}',t')}.
\end{equation}
The linear response function is also represented in Lehmann representation:\cite{L01}
\begin{equation}
\label{eqn:bkgnd:lehmann}
\tilde{\chi}(\vect{r},\vect{r}';\omega)=\lim_{\eta\to0_+}\sum_\alpha\left\{\frac{g_\alpha(\vect{r})g_\alpha^*(\vect{r}')}
{\omega-\Omega_\alpha+i\eta}-\frac{g_\alpha^*(\vect{r})g_\alpha(\vect{r}')}{\omega+\Omega_\alpha+i\eta}\right\},
\end{equation}
where $g_\alpha(\vect{r})=\matelemdirect{\Psi_\text{gs}}{\hat{n}(\vect{r})}{\Psi_\alpha}$ and $\Omega_\alpha=E_\alpha-
E_\text{gs}$.

Sum rules are moments of the oscillator-strength spectrum, and they are related to various theoretical or experimental 
physical properties of the ground-state atom. They are expressed with the following formula:\cite{BS57,FC68}
\begin{equation}
\label{eqn:sumrule:def}
S_j=\sum_s\omega_{s}^j \sigma_{s}+\int_I^\infty \intd\omega\: \omega^j \sigma(\omega),
\end{equation}
where $s$ denotes the discreet $1s\to np$ transitions and $j$ is an integer. We only use $-2\le j \le 2$ in this article. 
These sum rules have simple relations to physical properties, such as the ground-state density, polarizability, and 
kinetic energy, and thus they are easily calculated or determined from experiment. The specific relations we use are:
\begin{equation}
\label{eqn:sumrule:formula}
\begin{split}
&S_{-2}=\alpha(0),\quad S_{-1}=\frac{2}{3}\expvaluedirect{\absdirect{\textstyle\sum_j \vect{r_j}}^2}_0,\displaystyle\quad 
S_0=N,\\
&S_1=\frac{2}{3}\expvaluedirect{\absdirect{\textstyle\sum_j \vect{p_j}}^2}_0,\displaystyle\quad S_2=\frac{4}{3}\pi Z n(0),
\end{split}
\end{equation}
where $Z$ is the nuclear charge.
Eqn. \parref{eqn:sumrule:def} and Eqns. \parref{eqn:sumrule:formula} not only provide connections between the 
spectrum and several physical properties, but also imply that $S_0$, and $S_2$ are identical in the KS and the real 
spectrum, since the ground-state density in exact DFT is by definition equal to that of the real system. These equations 
also suggest the possibility of a fit which takes general physical properties as input and is able to generate the entire 
spectrum for the H atom.

As shown in Eqn. \parref{eqn:sumrule:def}, sum rules of the bound region of the spectrum are calculated by a 
summation of the discrete Rydberg states, and there is no trivial formula for calculating the energies of these states in 
multi-electron atoms. In order to characterize these energies for the summation, we use quantum defect theory\cite
{F06}. In quantum defect theory, the energy of the orbital with principle quantum number n in a multi-electron atom is 
expressed thus:
\begin{equation}
\label{eqn:bkgnd:qddef}
E_n=-\frac{1}{2(n-\mu_n)^2}.
\end{equation} 
This expression is used in calculating the bound part of sum rules if the formula for $\mu$ is known. The quantum defect 
is a smooth function of energy, and can be very accurately approximated\cite{FB06} by its Taylor expansion around $
\mu=0$:
\begin{equation}
\label{eqn:qd:formula}
\mu^{(p)}(E)=\sum_{i=0}^p \mu_i E^i,\quad E=\omega-I.
\end{equation}
For Helium, this curve is essentially linear, so $\mu\simeq\mu_0+\mu_1 E$, where $\mu_0=0.0164$ and $
\mu_1=0.0289$ for KS helium, and $\mu_0=-0.0122$ and $\mu_1=-0.0227$ for real helium. Inserting this expression 
into the $E_n$ formula and solving self-consistently yields highly accurate excitation energies\cite{FB09}

\begin{figure}[htbp]
\centering
\includegraphics[angle=-90,keepaspectratio]{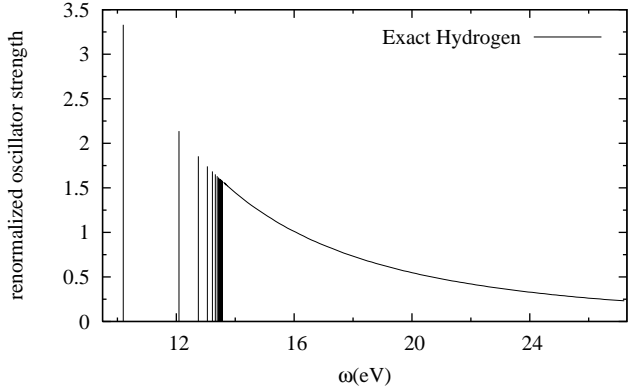}
\caption{Exact hydrogen spectrum. The ionization threshold is at 0.5 Hartree. Note that the bound-region spectrum has 
been renormalized so that it joins smoothly with the continuum-region spectrum.}
\label{fig:hydrogen}
\end{figure}

To illustrate these features in an exactly soluable case, we use the hydrogen spectrum as an example in this article. The 
exact form of the hydrogen oscillator strength is available using Eqn. \parref{eqn:bkgnd:fdef}:($n$ is the principal 
quantum number)
\begin{equation}
\label{eqn:bkgnd:hydrogenicf:bound}
\tilde{\sigma}_{1s\rightarrow np}=256n^8\left(\frac{n+1}{n-1}\right)^{-2n}/\left[3(n^2-1)^4\right],
\end{equation}
\begin{equation}
\label{eqn:bkgnd:hydrogenicf:unbound}
\tilde{\sigma}_{1s\rightarrow kp}=128\exp[h(k)]\csch\left(\frac{\pi}{k}\right)/\left[3(1+k^2)^4\right],
\end{equation}
where $h(k)=\left\{\pi+2\tan^{-1}\left[2k/(k^2-1)\right]-2\pi\theta(k-1)\right\}/2$, $k=\sqrt{2\epsilon}$ is the wavevector of 
the continuum wavefunction, and $\theta$ is the Heaviside step function. The wavefunctions of the bound states are 
energy-normalized by $n^{3/2}$, so the bound state wavefunctions and continuum wavefunctions agree with each other 
at ionization threshold. The hydrogen spectrum is shown in Fig. \ref{fig:hydrogen}. We represent the bound transitions 
as by simple line segments whose height is $\tilde{\sigma}$.

\section{High-frequency limit}
\label{sect:theory}

Fig. \ref{fig:thres} suggests the KS oscillator strength and exact oscillator strength share the same asymptotic form. Real 
oscillator strength spectra of atoms decay as $\omega^{-7/2}$\cite{RF67,KS57,FC68}. Here we derive the decay of the 
KS oscillator strength.

The oscillator strength is related to the transition dipole matrix element $\matelemdirect{\Psi_\text{f}}{r\cos\theta}{\Psi_
\text{i}}$ by Eqn. \parref{eqn:bkgnd:fdef}. In the KS system, the matrix element is greatly simplified, and can be written 
with one-electron KS orbitals as $\matelemdirect{\psi_\text{f}}{r\cos\theta}{\psi_\text{i}}$. For the absorption spectrum of 
the KS helium atom, the final orbital is a $p$ orbital with wavevector $\vect{k}$, and the initial orbital is the $1s$ orbital. 
In the high frequency limit of the absorption spectrum($\omega\to\infty$), $k\to\infty$ as well, and $\phi_\text{kp}(r)$ is 
highly oscillatory, where $\phi$ denotes radial wavefunctions. Then the matrix element is determined by the integrand 
near the nucleus. Thus the matrix element can be evaluated with the approximation of the initial KS orbital below:

\begin{equation}
\label{eqn:KS:approx}
\begin{split}
\phi_\text{i}(r)&=\exp(-\alpha r)[\phi_\text{i}(r)\exp(\alpha r)]\\
				&\approx\exp(-\alpha r)\left\{\phi_\text{i}'(0)r+\frac{1}{2}[2\alpha\phi_\text{i}'(0)+\phi_\text{i}''(0)]r^2
\right\},
\end{split}
\end{equation}
where $\phi_{i}$ is the spherical wavefunction of the initial KS orbital, and $\alpha$ is a positive real number 
characterizing the decay of the wavefunction. The cusp condition\cite{K57} holds in KS helium, so $\phi_\text{i}''(0)=-2Z
\phi_\text{i}'(0)$, where $Z=2$ is the nucleus charge. Then $\phi_\text{i}$ is rewritten as

\begin{equation}
\label{eqn:KS:phiinit}
\phi_\text{i}(r)\approx\exp(-\alpha r)\{r+(\alpha-Z)r^2\}\phi_\text{i}'(0).
\end{equation}

In $k\to\infty$ limit, only the $-2/r$ Coulomb well in the KS potential is important to $\phi_\text{kp}$. Then $\phi_\text{kp}$ 
is approximated with hydrogenic wavefunctions, and the approximation becomes exact when $k\to\infty$. The transition 
dipole matrix element is evaluated at $k\to\infty$ limit.

\begin{equation}
\label{eqn:KS:matelem}
\matelemdirect{\psi_\text{f}}{r\cos\theta}{\psi_\text{i}}\:\to\:\left(4\sqrt{\frac{2}{3\pi}}Z\phi_\text{i}'(0)\right)k^{-9/2}\quad,\,k\to
\infty.
\end{equation}

The oscillator strength spectrum then decays as
\begin{equation}
\label{eqn:KS:oscstrdecay}
\tilde{\sigma}(\omega)\:\to\:\frac{2\sqrt{2}}{3\pi}[\phi_\text{i}'(0)Z]^2\omega^{-\frac{7}{2}}\quad,\,\omega\to\infty.
\end{equation}

Eqn. \parref{eqn:KS:oscstrdecay} implies the asymptotic decay of the oscillator strength only depend on the properties 
at the nucleus. For hydrogen and helium, Eqn. \parref{eqn:KS:oscstrdecay} is related to the electronic density by
\begin{equation}
\label{eqn:AsymForm:asym}
\tilde{\sigma}(\omega)\:\to\:\frac{8\sqrt{2}}{3}Z^2n(0)\omega^{-\frac{7}{2}}\quad,\,\omega\to\infty.
\end{equation}

For hydrogen, the coefficient of the $\omega^{-7/2}$ term is $8\sqrt{2}/3\pi$. Eqn. \parref{eqn:KS:oscstrdecay} and 
\parref{eqn:AsymForm:asym} give the correct result. With these equations, the asymptotic behavior of the oscillator 
strength spectrum is determined. The discussion of the high-frequency part of the KS helium oscillator strength 
spectrum can be extended to other KS atoms easily, as the KS system is an one-electron picture. Following similar 
procedure as described here, it can be easily verified that the high-frequency part of the KS oscillator strength of other 
KS atoms can also be expressed in terms of the density at the nucleus. As only $s$ orbitals has non-zero contribution to 
the density at the nucleus, one would expect that Eqn. \parref{eqn:AsymForm:asym} also holds for other atoms, using 
the corresponding $Z$ and $n(0)$.

\begin{figure}[htbp]
\centering
\includegraphics[angle=-90,keepaspectratio]{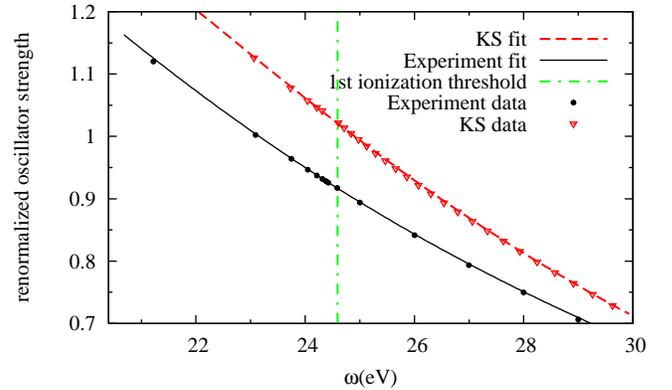}
\caption{KS and experimental single-electron oscillator strength and fit curve of He near threshold.\cite
{SHYH94,WMB03} The ionization threshold is at 0.9036 Hartree. The curves are converted from $g(x)$ fit(Eqn. \parref
{eqn:gx}). The upper curve represents the exact KS Helium oscillator strength data and fit curve, and the lower curve 
represents the experimental Helium oscillator strength data and fit curve.}
\label{fig:HeGx}
\end{figure}

The half-power decay of Eq. (23) differs noticeably from the decay discussed by van Leeuwen, but here we are 
considering the imaginary part of the response function, whereas there it is the real part of the response function. We 
are currently investigating relationship between the two in the general case.

\section{Fitting of 1- and 2-electrons spectra using sum rules}
\label{sect:fitting}

We fit the oscillator strength spectra to answer the title question. Since we want to study the near-threshold behavior of 
the oscillator strength spectrum, the position of the ionization threshold is treated explicitly in our fit. We define $x$ and 
$g(x)$ as

\begin{equation}
\label{eqn:defgx}
\begin{split}
x=2(\omega-I),\\
g(x)=\frac{3\omega^4}{8}\tilde{\sigma}(\omega).
\end{split}
\end{equation}

The fit has to satisfy a few criteria to generate the correct shape for the oscillator strength spectrum. The fit is employed 
to study the exactness of the KS oscillator strength at the ionization threshold, where the fit needs to have the correct 
series expansion. We take the expansion of the hydrogen oscillator strength at the ionization threshold:

\begin{equation}
\label{eqn:fit:seriesI}
\tilde{\sigma}(\omega\to I)=c_0+c_1(\omega-I)+c_2(\omega-I)^2+\cdots.
\end{equation}

We assume the fit formula has the same expansion near the ionization threshold. This assumption is justified by the 
following consideration. Near the ionization threshold, the oscillator strength spectrum of real helium is determined by 
the Rydberg states, which resembles the hydrogenic states. The KS helium is a system with non-interacting electrons, 
so the oscillator strength spectrum resembles that of one-electron systems.

We use the fit to show that the oscillator strength spectrum around the ionization threshold is smooth, that no near-
degeneracies exist around the ionization threshold. Thus the fit also need to accurately generate the entire oscillator 
strength spectrum, including both discrete and continuum regions, so that the conclusions from the fit are convincing. To 
generate the correct continuum spectrum, the fit need to have the correct series expansion when $\omega\to\infty$. As 
in Sect. \ref{sect:theory}, the asymptotic series expansion of helium has the same form as hydrogen:

\begin{equation}
\label{eqn:fit:seriesasymp}
\tilde{\sigma}(\omega\to\infty)=d_1\omega^{-\frac{7}{2}}+d_2\omega^{-4}+d_3\omega^{-\frac{9}{2}}+\cdots.
\end{equation}

\begin{figure}[htbp]
\centering
\includegraphics[angle=-90,keepaspectratio]{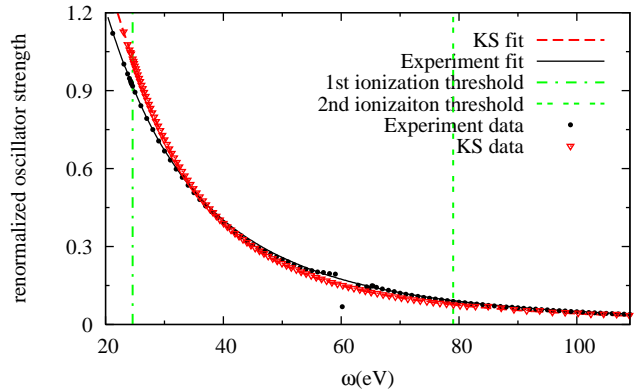}
\caption{KS and experimental single-electron oscillator strength and fit curve of He.\cite{SHYH94,WMB03} This figure 
show the overall shape of the oscillator strength curves. The solid dots and curve represents the exact KS Helium 
oscillator strength data and fit curve, and the cross dots and dashed curve represents the experimental Helium oscillator 
strength data and fit curve.}
\label{fig:HeBig}
\end{figure}

\begin{figure}[htbp]
\centering
\includegraphics[angle=-90,keepaspectratio]{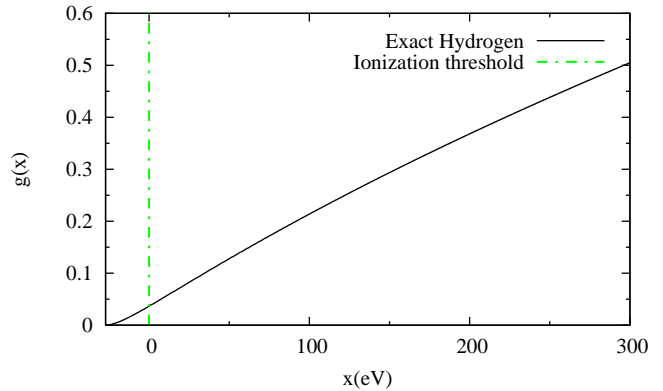}
\caption{$g(x)$ of Hydrogen(Eqn. \parref{eqn:defgx}). The ionization threshold is at $x=0$.}
\label{fig:Hgx}
\end{figure}

The shape of the $g(x)$ function is shown in Fig. \ref{fig:Hgx}.  Our $g$-fit formula is
\begin{equation}
\label{eqn:gx}
g(x)=a+b[1-\exp(-cx)]+d\sqrt{e+x},
\end{equation}
where $a$, $b$, $c$, $d$, $e$ are fit parameters. Note that aside from giving the correct series expansion at the 
ionization threshold and asymptotically, the form does not have other explicit physical motivation. It is solely designed to 
recover the shape of the oscillator strength curves. We determine the parameters by the process below.

The important points are fixed on the fit. Since we use the fit to study the oscillator strength around the ionization 
threshold, we fix the value and the first derivative of the oscillator strength at the ionization threshold. The asymptotic 
coefficient in Sect. \ref{sect:theory} is not fixed, but used as the initial point of search. The remaining three parameters 
are determined by applying oscillator strength sum rules(Eqn. \parref{eqn:sumrule:def}) to the fit curve. We evaluate the 
sum rule of a fit curve by adding the contributions from the discrete transitions and that from the continuum. For the 
discrete region, we calculate the frequency of a transition with quantum defect theory(refer to Eqn. \parref
{eqn:bkgnd:qddef} and \parref{eqn:qd:formula}). The oscillator strength of the transition is then evaluated with the fit 
formula(with a certain initial choice of parameters). We add the contribution of different discrete transitions up to 
$n=1000$. For the continuum region, we carry out a numerical integration over the entire continuum.

The exact values of the oscillator strength sum rules are available for both KS and real helium, since these sums are 
related to various physical properties (refer to Eqn. \parref{eqn:sumrule:formula}). To fit the oscillator strength spectrum, 
we choose an initial set of the parameters. Only three fit parameters are independent, so we choose three sum rules to 
fit. We minimize the difference between the sums evaluated on the fit curves and the exact sums obtained from physical 
properties by varying the three parameters numerically. The search ends when the accuracy of the fitted sums reach a 
predetermined goal. In our application, the difference between the sums of the fit and the exact sums is smaller than 
$10^{-8}$. The accuracy of the fit is also checked by evaluating the unused sum rules (Table. \ref{tab:rescomp}).

The fit can use two to four sum rules depending on how many points are fixed in the beginning. Applying more sum 
rules increases the overall accuracy of the fit, but the process of numerically fitting sum rules becomes more difficult. All 
results in this paper are obtained with three sum rules. With Eqn. \parref{eqn:gx}, the sum rules of the fit curve can be 
written out in terms of the parameters.
\begin{equation}
\label{eqn:gxsum}
\begin{split}
S_j&=S_j^\text{dis}+\frac{8}{3}\left\{2^{3-j}d\left(2I-e\right)^{j-5/2}B_{1-e/2I}\left(\frac{5}{2}-j,\frac{3}{2}\right)\right.\\
&\quad\left.+I^{j-3}[a+b+b\left(j-3\right)\exp\left(2cI\right)E_{4-j}\left(2cI\right)]/(3-j)\right\},\\
S_j^\text{dis}&=\frac{8}{3}\sum_{n=2}^\infty\beta^{3}\left\{a+b\left[1-\exp\left(c\beta^2\right)\right]+d\sqrt{e-\beta^2}\right\}
\gamma.
\end{split}
\end{equation}
where $\beta=\left(-\mu_0-\mu_1/n^2+n\right)^{-1}$, $\gamma=\left(I-\beta^{2}/2\right)^{-4+j}$, $\mu_0$ and $\mu_1$ 
are the parameters in the quantum defect forumla(Eqn. \parref{eqn:qd:formula}), $B$ is the incomplete beta function, 
and $E$ is the exponential integral function\cite{AS72}.

\begin{figure}[htbp]
\centering
\includegraphics[angle=-90,keepaspectratio]{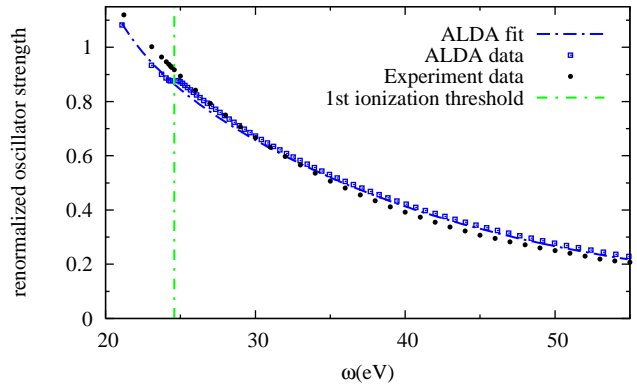}
\caption{Exact/ALDA oscillator strength and fit curve of He. These oscillator strength data are obtained from an ALDA 
calculation with exact KS ground state. We use a box code\cite{FB09} to calculate these data. There is a kink in our data 
near the ionization threshold, because the continuum near the ionization threshold mixes with higher Rydberg states, 
which are not well-described by the box code.}
\label{fig:HeExactALDA}
\end{figure}

With Eqn. \parref{eqn:gx}, we obtain the oscillator strength curves of KS Helium and real Helium. We also apply our 
method to the ALDA Helium(with exact KS ground state) as the first step of studying the threshold behavior in TDDFT
(Fig. \ref{fig:HeExactALDA}). The comparison of results and figures of oscillator strength curves are shown in Sect. \ref
{sect:results} and in Fig. \ref{fig:HeGx}.

Note that the fit is not designed to be used as an interpretation tool, but to recover the shape of the oscillator strength 
spectrum. Thus comparing the fit parameters of different curves(exact KS, ALDA, and experimental) is largely 
meaningless as there is no visible trend. An exception is the fit parameter $d$, which describes the shape of the 
asymptotic part of the oscillator strength curve, as it is related to the coefficient of the leading term($\omega^{-7/2}$) of 
the asymptotic expansion of the oscillator strength. The fit parameters of related systems are provided in the
supplementary material.\cite{SupplMat}

\begin{table}[tbp]
\centering
\caption{Sum rules from $\tilde{\sigma}(\omega)$ and $g(x)$ fit}
\label{tab:rescomp}
\begin{minipage}[t]{0.9\columnwidth}
\renewcommand{\thefootnote}{\alph{footnote}}
\hfill
\begin{tabular}{llcccc}
\hline
														&			&	$S_{-2}$	&	$S_{-1}$	
&	$S_1$	&	$S_2$\\
\hline\hline
														& g-fit &	4.4999		&	2\footnotemark
[3]&	0.6667&	1.3371\\
\raisebox{2ex}[0pt]{H}										& exact		&	4.5			&	2	
		&	2/3		&	4/3	\\
\hline
								& g-fit &	0.7563	&	0.7952	&	1.9114\footnotemark[3]	&15.167
\footnotemark[3]\\
\raisebox{2ex}[0pt]{He KS\footnotemark[1]}						& exact		&	0.7579\cite{GKSG98}&	
0.7957\footnotemark[4]	&	1.9114\cite{HU97}&15.167\footnotemark[4]\\
\hline
								&g-fit	&	0.691	&	0.7504	&	2.09\footnotemark[3]	&	15.167
\footnotemark[3]\\
\raisebox{2ex}[0pt]{He Exp.\footnotemark[1]\footnotemark[2]}			& exact		&	0.698		&	
0.754		&	2.09	&	15.167\footnotemark[4]\\
\hline
														& g-fit &	0.6912\footnotemark[3]	&	
0.7519	&	2.0414	&	15.167\footnotemark[3]\\
\raisebox{2ex}[0pt]{He ALDA\footnotemark[1]}			& exact		&	0.6912\cite{GKSG98}&	0.7957	&	
1.9114\cite{HU97}&15.167\\
\hline
\end{tabular}
\hfill
\footnotetext[1]{All the sums are converted to corresponding single-electron sums}
\footnotetext[2]{The expected value of experimental data are listed in \cite{FC68}}
\footnotetext[3]{This sum rule is a constraint. $S_0$ is always a constraint.($S_0=1$ for all systems after converted to 
single-electron model)}
\footnotetext[4]{The expected value of $S_{-1}$ and $S_2$ are calculated from the exact helium density\cite{UG94}}
\end{minipage}
\end{table}

\section{Results}
\label{sect:results}

\begin{figure}[htbp]
\centering
\includegraphics[angle=-90,keepaspectratio]{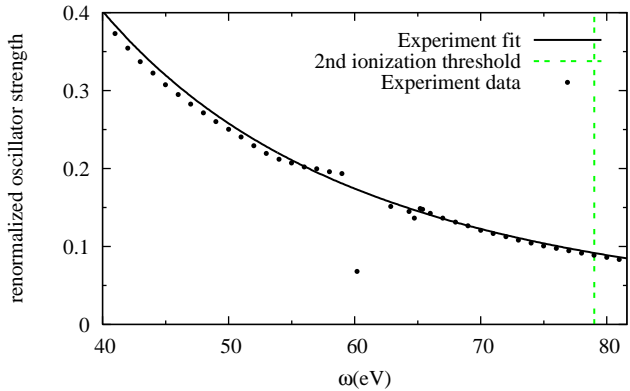}
\caption{$g$-fit of the experimental helium oscillator strengths near the autoionizing resonances.}
\label{fig:HeResonance}
\end{figure}

The $g$-fit curves of the KS helium and the real helium are shown in Fig. \ref{fig:HeGx} and \ref{fig:HeBig}. The fit is very 
accurate in the entire range of $\omega$. The accuracy is also checked with the unused sum rules, listed in Table. \ref
{tab:rescomp}. The results of the hydrogen atom are listed as a reference, and it shows that the inherent error of the 
method is small. With these curves, we explicitly show that the oscillator strength spectrum is a smooth curve around the 
ionization threshold, and thus the oscillator strength of the exact KS helium is not that of the real helium. The 
autoionizing resonances in real helium are not included in our fit, but the fit is still accurate even near the resonances
(Fig. \ref{fig:HeResonance}). The errors in the sum rules are small, so the fit curve can be used as a background for 
studying these autoionizing resonances, and the pure resonance peaks can be obtained by subtracting the fit curve 
from the experimental spectrum.

One reason for the good performance near the resonances is that the autoionization resonances occur at relatively high 
frequencies, so their contributions to the smaller sum rules are neglectable. The other reason is the shape of the 
autoionization resonances in He is asymmetric, which have both a dip and a peak in the resonance region.\cite
{F61,FI90} The contribution of these two parts to the sum rules cancels, so the values of the sum rules are not influenced 
by the autoionization resonances too much(even for $S_2$), and thus the fit accurately generates the oscillator strength 
curves for He.

\section{Conclusions}
\label{sect:conclusion}

If the answer to the title question had been yes, then it would yield a strong exact condition on the XC kernel in TDDFT,
 which many approximations would fail. Thus we studied the title problem. We have shown that Kohn-Sham oscillator 
 strength of He is not exact at the ionization threshold (even though the position of the threshold is exact), and so the 
 answer to the title question is no. This implies that the Hartree-XC kernel in TDDFT has non-zero off-diagonal matrix 
 elements at the threshold, and simple approximations such as the single-pole approximation are insufficient in this 
 region.

We also developed a numerical fit to generate the spectrum near the ionization threshold from a few physical conditions 
such as sum rules. The fit is accurate for all frequencies due to the smoothness of the oscillator strength near the
 threshold, but also works well for the spectrum far from the ionization threshold due to the correct asymptotic behavior. 
The fit is not physically motivated, but is a simple accurate representation of the curves.

These results are not general since we only studied atoms with one or two electrons, and multi-electron resonances are 
ignored as in Fig. \ref{fig:HeGx}. However, obvious generalizations can be performed for atoms with more electrons 
since we only use the general properties(the asymptotic behavior, the value and first derivative of the spectrum at 
ionization threshold, and sum rules) in our method. Thus multi-electron resonances can be dealt with by subtracting 
their contribution from sum rules, and thus our method can be extended to other atoms by following the methods of Sect. 
\ref{sect:theory} and \ref{sect:fitting}.

\section*{Acknowledgements}
We thank Cyrus Umrigar for providing us with his exact Kohn-Sham potentials for the Helium atom, Robert van 
Leeuwen for helpful discussions, and Adam Wasserman for his oscillator strength data. This work is funded by the U.S. 
Department of Energy(Grant No. DE-FG02-08ER46496).


\end{document}